\documentclass[journal]{IEEEtran}

\ifCLASSINFOpdf
\else
\fi

\usepackage{graphicx}
\usepackage{amssymb}
\usepackage{pifont}
\newcommand{\xmark}{\ding{55}}%
\usepackage{amsmath}
\usepackage{multirow}
\usepackage{xcolor}
\usepackage[ruled,linesnumbered,noresetcount]{algorithm2e}
\begin{document}
%
\title{Self-Supervised Learning based Monaural Speech Enhancement with Complex-Cycle-Consistent}
%
%
%

\author{Yi~Li,~\IEEEmembership{Student Member,~IEEE,}
        Yang~Sun,~\IEEEmembership{Member,~IEEE,}
        and~Syed~Mohsen~Naqvi,~\IEEEmembership{Senior~Member,~IEEE}
\thanks{Y. Li and S. M. Naqvi are with the Intelligent Sensing and Communications Group, School of Engineering, Newcastle University, Newcastle upon Tyne NE1 7RU, U.K. (e-mails: y.li140, mohsen.naqvi@newcastle.ac.uk)
}
\thanks{Y. Sun is working with the Big Data Institute, University of Oxford, Oxford OX3 7LF, U.K. (e-mail: Yang.sun@bdi.ox.ac.uk)}

\thanks{E-mail for correspondence: y.li140@newcastle.ac.uk}
}
\markboth{Journal of \LaTeX\ Class Files,~Vol.~14, No.~8, August~2015}%
{Shell \MakeLowercase{\textit{et al.}}: Bare Demo of IEEEtran.cls for IEEE Journals}
%



\maketitle

\begin{abstract}
Recently, self-supervised learning (SSL) techniques have been introduced to solve the monaural speech enhancement problem. Due to the lack of using clean phase information, the enhancement performance is limited in most SSL methods. Therefore, in this paper, we propose a phase-aware self-supervised learning based monaural speech enhancement method. The latent representations of both amplitude and phase are studied in two decoders of the foundation autoencoder (FAE) with only a limited set of clean speech signals independently. Then, the downstream autoencoder (DAE) learns a shared latent space between the clean speech and mixture representations with a large number of unseen mixtures. A complex-cycle-consistent (CCC) mechanism is proposed to minimize the reconstruction loss between the amplitude and phase domains. Besides, it is noticed that if the speech features are extracted as the multi-resolution spectra, the desired information distributed in spectra of different scales can be studied to further boost the performance. The NOISEX and DAPS corpora are used to generate mixtures with different interferences to evaluate the efficacy of the proposed method. It is highlighted that the clean speech and mixtures fed in FAE and DAE are not paired. Both ablation and comparison experimental results show that the proposed method clearly outperforms the state-of-the-art approaches.
\end{abstract}


%
\IEEEpeerreviewmaketitle

\section{INTRODUCTION}
\IEEEPARstart{I}{n} recent years, deep learning techniques have significantly improved the speech enhancement performance in a wide range of real-world applications such as assisted living systems, teleconferencing, and automatic speech recognition (ASR) \cite{yiluo, asr}. However, the novel networks are predominantly trained in a supervised mechanism where requires a vast set of paired data as clean speech signals and the corresponding mixtures. To exploit the models in highly reverberant scenarios, in self-supervised learning (SSL), the relationships and similarities between the training samples are applied to estimate the corresponding paired labels for the training set \cite{ssl6}.

Recently, self-supervised techniques have been applied in speech enhancement problem. Wang et al. use an autoencoder to learn a latent representation of clean speech signals and autoencode on speech mixture with the shared representation of the clean examples \cite{ssl}. However, the pre-training stage only comprises one pre-task which maps the amplitude of the mixture spectrogram to the clean speech. To solve the insufficient pre-training limitation, Du et al. propose the self-supervised adversarial learning to improve the noise generalization ability \cite{ssl2}. The autoencoder obtains both magnitude and phase feature information in the complex spectrum and benefits speech enhancement. In speech enhancement study, the encoder takes a spectrogram as the input and transforms it into a latent space and the decoder maps the encoded representation to the original-size spectrogram. Therefore, the decoder plays an important role in recovering the clean speech from the latent space. However, the performance is limited when the magnitude and phase components share the same decoder due to the reconstruction loss of one component caused by the other component. Different from the conventional complex spectrogram methods, we apply two individual decoders in the foundation autoencoder (FAE) to process the magnitude and phase information as presented in Fig. 1. Then, two decoders are used to produce the estimated magnitude and phase of the target speech signal. Moreover, we provide the comparisons with two encoders autoencoder which helps to set configurations in the pipeline. The experimental results are shown in Section III.
\begin{figure*}[htbp!]
\centering
\includegraphics[width=16cm, height=8.5cm]{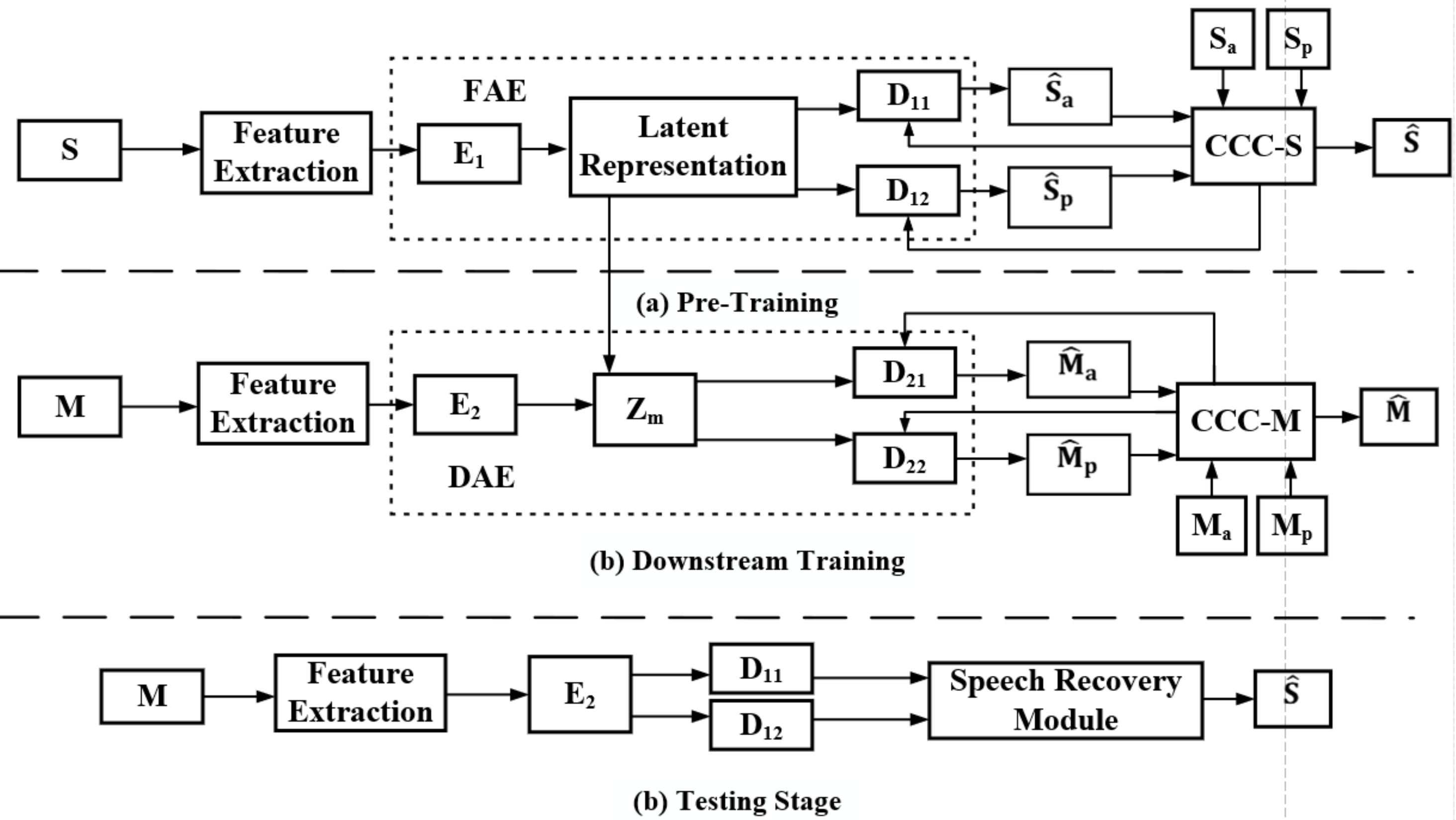}
\caption{The overall architecture of the proposed method. Generally, the foundation autoencoder (FAE) consists of $E_{1}$, $D_{11}$, and $D_{12}$, while downstream task autoencoder (DAE) is composed of $E_{2}$, $D_{21}$, and $D_{22}$. As the input of $E_{1}$, the multi-resolution features are extracted from the clean spectra with different window sizes. Then, the latent representation of the clean speech signal is learned via $E_{1}$ and $E_{2}$. After the multi-resolution feature maps are recovered in the decoder, the reconstructed clean spectra are obtained as the output by using $D_{11}$ and $D_{12}$. Besides, the reconstructed and target spectra are fed into the complex-cycle-consistent for speech module to further train the FAE. The estimated mixtures are produced from the downstream task autoencoder (DAE) which shares the learned representation. The mixture latent representation is presented as $Z_{m}$. In the testing stage, the enhanced signal is obtained with the estimated spectrogram from the speech recovery module. }\centering
\end{figure*}

The consistent learning-based methods achieve great success in speech enhancement to minimize the reconstruction loss. For example, Meng et al. propose a cycle-consistent method to speech enhancement problem \cite{csse}, a clean-to-noisy mapping network is added to the noisy-to-clean network to reconstruct the noisy features from the enhanced ones. Compared to the conventional CycleGAN-based methods, the target speech phase is estimated by a complex spectral network \cite{TCN}. In this work, we propose a complex-cycle-consistent (CCC) mechanism to calculate the commutative losses between the amplitude and phase features and utilize in SSL based speech enhancement problem for the first time.

Meanwhile, researchers exploit multiple resolutions or scales to improve the speech enhancement performance, where the features are extracted and weighted differently as inputs to the neural networks \cite{multis}. For example, convolutional blocks with different scales are used to process the same input time-frequency (TF) features for the end-to-end automatic speech recognition \cite{twos}. A channel-aware attention mechanism is introduced to enforce the connections between feature groups in neural networks. The attention mechanism is combined with convolutional neural networks (CNN) where the key features of anti-spoofing are explored by assigning weights to different positions and channels in a feature map \cite{icme}. However, a vast training set of paired examples of clean speech signals and the corresponding mixtures are required at the training stage \cite{ssl}. In this work, we propose the first SSL work to address speech enhancement problem with multi-resolution spectra.

The contributions of the paper are summarized as follows:

$\bullet$ We use two independent decoders to learn the latent representations of both amplitude- and phase-related features. 

$\bullet$ Although phase-based methods are commonly used in speech enhancement methods, based on the phase feature of the spectrogram, we can further utilize the CCC mechanism between the amplitude and phase feature maps to minimize the reconstruction loss for each other. To the best of our knowledge, it is the first time that the cycle-consistent approach is applied in SSL based speech enhancement problem.

$\bullet$ The multi-resolution spectra losses are also introduced in the proposed phase-aware SSL enhancement method to further improve the speech enhancement performance.

\section{RELATED WORK}
With a view to relax the constraints of paired training data, in speech enhancement problem, many corresponding approaches are proposed. For example, in weakly supervised approaches, rather than using representative clean training examples to extract the target speech signal, the techniques use various weakly supervised labels. Kong et al. confirm that weakly supervised training in combination with supervised training improves performance over standalone supervised training \cite{weak}. Moreover, semi-supervised learning combines both labeled and unlabeled data with pseudo-labels (PLs) generated in one way or another \cite{poco, semi}. However, according to \cite{ssl}, these methods cannot be reused in unmatched training and testing conditions. 

Inspired by the original research in image-to-image translation, the CycleGAN is introduced in speech enhancement problem \cite{csse}. Exploiting CycleGAN in speech enhancement problem, the state-of-the-art methods show the effectiveness in improving the performance particularly in reducing noise interferences and remaining speech integrity \cite{TCN}. Generally, the CycleGAN consists of a noisy-to-clean generator $G$ and an inverse clean-to-noisy generator $F$, which transforms the noisy features into the enhanced ones for the former, and vice versa for the latter. A forward noisy-clean-noisy cycle and a backward clean-noisy-clean cycle jointly constrain $G$ and $F$ to be cycle-consistent, which are optimized with the adversarial loss, a cycle-consistency loss, and an identity-mapping loss, respectively \cite{csse}. Discriminators are trained to classify the target speech features as real and the generated speech features as fake. Although it achieves compelling results in the experiments, the reconstruction is far from the clean speech signals.
\begin{table*}[htbp!]
\caption{speech enhancement performance at three SNR levels (-5, 0, and 5 dB) in ipad$\char`_$bedroom1 environment.}
\centering
\begin{tabular}{ccccccccccccc}
\hline
\multirow{2}{*}{Model} & \multicolumn{3}{c}{PESQ} & %
    \multicolumn{3}{c}{CSIG} & \multicolumn{3}{c}{CBAK}& \multicolumn{3}{c}{COVL} \\
\cline{2-13}
 &-5 dB& 0 dB& 5 dB&-5 dB& 0 dB& 5 dB&-5 dB& 0 dB& 5 dB&-5 dB& 0 dB& 5 dB\\
 SSE \cite{ssl}  &1.59&1.62&1.65 &2.34&2.43&2.49 & 1.88&1.97&2.16&1.84&1.89&2.02 \\
\textit{Proposed}   &{\bfseries 1.88}&1.92&1.95 &{\bfseries 2.49}&{\bfseries 2.56}& 2.61 & {\bfseries 2.02}&  2.16& {\bfseries 2.25}& {\bfseries 1.96}&  2.05&  2.13\\

Two Encoders  &1.87&{\bfseries 1.94}&{\bfseries 1.98} &2.45&{\bfseries 2.56}&{\bfseries 2.62} & 2.00&{\bfseries 2.19}&{\bfseries 2.25}&1.93&{\bfseries 2.09}&{\bfseries 2.15} \\
 \hline 
\end{tabular}
\end{table*}

In speech enhancement problem, the Short-time Fourier transform (STFT) is widely used \cite{ssl7, CSA1, mcgn} to achieve feature extraction. The main idea is supported by the recent research in speech processing that it is not clear what cues in which scales of window lengths contribute most to the final performance \cite{multis}. Thus, the feature is extracted from a combined input of multi-resolution feature maps to fully use the desired information with different scales. It has been confirmed that the multi-resolution features have a higher time and frequency domain resolution structure of speech than the single-resolution \cite{mr}. Therefore, it is expected that the multi-resolution can capture the temporal dynamics of emotion cues in natural speech to improve the prediction accuracy \cite{pwgan}. In most cases, it is challenging to promise that the input single-resolution spectrogram bring the best performance compared with other resolutions \cite{vogan}. Therefore, in the proposed work, the feature is extracted from multi-resolution spectra and a combined loss of spectra is minimized to better use the desired information on feature maps with different resolutions.

\section{PROPOSED METHOD}
The overall model architecture of the proposed phase-aware multi-resolution autoencoder is presented in Fig. 1. In this paper, we adopt the variational autoencoder (VAE) as the primary framework for two reasons. First, the generative adversarial network (GAN) and its varieties suffer a limitation as the difficulty of model training due to destabilization \cite{ganno1}. Second, in the GAN based methods, unseen data may be mapped out of the subspace, leading to poor results \cite{ganno}. Particularly within SSL cases where a limited training set of labelled data is applied, as proved in \cite{ganno2}, the VAE performs better at learning the low-dimensional latent space. In the training stage, two variational autoencoders \cite{VCAE, TAI}, foundation autoencoder (FAE) and downstream task autoencoder (DAE), are exploited for different tasks. In order to extract the amplitude and phase from the feature map, we temporarily set the number of encoders in each autoencoder to one and two, and compare the speech enhancement performance in TABLE I. It can be observed that the speech enhancement performance of the two encoders baseline is limited compared to the proposed single encoder method in some SNR levels. Although the performance is improved with the two encoders baseline in some results such as at 0 dB in terms of PESQ, the improvement is limited and the training computational cost increases due to the second encoder. Therefore, we only use one encoder in each FAE and DAE to extract the amplitude and phase jointly. The encoders are denoted as $E_{1}$ and $E_{2}$ in FAE and DAE, respectively. Second, in \cite{ssl}, each autoencoder is comprised of one encoder and one decoder. However, the proposed method exploits two decoders for amplitude and phase in each autoencoder. We define them as $D_{11}$ and $D_{12}$ for the amplitude and phase in FAE, respectively. Meanwhile, in DAE, $D_{21}$ and $D_{22}$ are trained to learn the amplitude and phase, respectively.

Initially, the spectra of a limited set of clean speech signals are obtained by using STFT as the input of FAE. The mel-frequency cepstral coefficients (MFCC) feature is extracted from the spectra. In order to better preserve the desired information distributed in multi-resolution feature maps, in the encoder $E_{1}$, each layer obtains one feature map and produce a latent representations of the clean speech signal. The feature maps are scaled as different resolutions. In the training of $E_{1}$, optimal weighted combinations of multi-resolution spectra are learned given the objective of the clean speech representation.

In the proposed method, we consider two pre-tasks in pre-training, one is used to learn the amplitude feature information and another aims to learn the phase of the latent representation. Therefore, two decoders $D_{11}$ and $D_{12}$ are applied to learn the amplitude and phase of clean speech, respectively. In details, both the amplitude and phase latent representations are learned by minimizing the discrepancy between the input representation and the corresponding reconstruction. The multi-resolution spectra of the estimated speech signal are obtained and compared with the clean spectra. At the same time, the combined loss of the spectra are used to train the FAE to extract the target speech signal.

In FAE, each $E_{1}$, $D_{11}$, and $D_{12}$ consists of 4 1-D convolutional layers. In $E_{1}$, the size of the hidden dimension decreases sequentially from 512 $\rightarrow$ 256 $\rightarrow$ 128 $\rightarrow$ 64. Accordingly, the dimension of the latent space is set to 64, and a stride of 1 sample with a kernel size of 7 for the convolutions. Different from $E_{1}$, $D_{11}$, and $D_{12}$ increase the size of the latent dimensions inversely.

Different from the FAE, the DAE only needs access to the speech mixture. The feature is extracted from the speech mixture and fed to $E_{2}$. Consequently, the latent representation of the mixture is obtained as the output of $E_{2}$ and exploited to modify the loss functions and learn a shared latent space between the clean speech and mixture representations. Benefited from the pre-tasks, a mapping from the mixture domain to the target speech domain is learned with the latent representation of the clean speech signal. Furthermore, $D_{21}$ and $D_{22}$ are trained to produce the amplitude and phase of the estimated mixture as the downstream task, respectively. 

The DAE network follows a similar architecture to FAE. $E_{2}$ consists of 6 1-D convolutional layers where the hidden layer sizes decrease from 512 $\rightarrow$ 400 $\rightarrow$ 300 $\rightarrow$ 200 $\rightarrow$ 100 $\rightarrow$ 64, and decoders increase the sizes inversely.

In the testing stage, once the trained $E_{2}$, $D_{11}$, and $D_{12}$ are obtained, the feature of the mixture is extracted and fed to the trained model. Finally, as the outputs of the two decoders, the phase is recovered by re-wrapping the estimated unwrapped phase of speech in the speech recovery module and used to produce the estimated signal with the recovered speech amplitude. 
\subsection{Phase-Aware Loss}
Conventionally, the polar coordinate representation of of the clean speech $\mathbf{S}$ and the mixture $\mathbf{M}$ are written as:
\begin{equation}
\mathbf{S}=\mathbf{S}_{a} \cdot e^{j \mathbf{S}_{p}}
\end{equation}
\begin{equation}
\mathbf{M}=\mathbf{M}_{a} \cdot e^{j \mathbf{M}_{p}}
\end{equation}
where $j = \sqrt{-1} $ and the subscripts ‘$a$’ and ‘$p$’ indicate the amplitude and the phase components, respectively. Phase is generally difficult to estimate, especially in time-frequency (T-F) units with low SNR levels \cite{letter}. In recent speech enhancement study, the speech signal is usually reconstructed by using the noisy phase and the estimated magnitude \cite{ssl5}. However, phase estimation plays a pivotal role in the target speech signal reconstruction as the significant difference between the phase of the clean speech and mixture. Therefore, the aim of the proposed phase-aware method is to minimize the loss between both the amplitude and phase of the clean speech signal and the corresponding reconstruction. The amplitude loss can be presented:
\begin{equation}
\mathcal{J}_{\text {S}_{a}}= \sum_{i=1}^{I}\|\mathbf{S}^{i}_{a}-\hat{\mathbf{S}}^{i}_{a}\|_{2}^{2}
\end{equation}
where $i$ denotes the number of the spectra resolution in the proposed multi-resolution method. According to \cite{ssl}, we use the $L$2 norm $\|\cdot\|_{2}^{2} $ to estimate the loss terms. As aforementioned, $E_{1}$ consists of 4 1-d convolutional layers and each layer obtains the spectra as one specific resolution. Therefore, $I$ is set to 4 as the maximum number of the spectra resolutions and $\mathbf{S}_{a}$ denotes the combination of four $\mathbf{S}^{i}_{a}$. Besides, the amplitude of the clean speech spectra and the reconstruction are showed as $\mathbf{S}^{i}_{a}$ and $\hat{\mathbf{S}}^{i}_{a}$, respectively. Similarly, the phase loss can be presented:
\begin{equation}
\mathcal{J}_{\text {S}_{p}}= \sum_{i=1}^{I}\|\mathbf{S}^{i}_{p}-\hat{\mathbf{S}}^{i}_{p}\|_{2}^{2}
\end{equation}
where $\mathbf{S}^{i}_{p}$ and $\hat{\mathbf{S}}^{i}_{p}$ are the phases of the clean speech signal and the reconstruction. Then, the clean spectrogram loss $\mathcal{J}_{\text {S}}$ is the addition of $\mathcal{J}_{\text {S}_{a}}$ and $\mathcal{J}_{\text {S}_{p}}$. By minimizing $\mathcal{J}_{\text {S}}$, FAE is trained to learn a latent representation as a zero-mean normal distribution.

Similar to the target speech, in order to estimate the discrepancy between the speech mixture and the corresponding reconstruction, we exploit two losses as:
\begin{equation}
\mathcal{J}_{\text {M}_{a}}= \sum_{i=1}^{I}\|\mathbf{M}^{i}_{a}-\hat{\mathbf{M}}^{i}_{a}\|_{2}^{2}
\end{equation}
\begin{equation}
\mathcal{J}_{\text {M}_{p}}= \sum_{i=1}^{I}\|\mathbf{M}^{i}_{p}-\hat{\mathbf{M}}^{i}_{p}\|_{2}^{2}
\end{equation}
where $\hat{\mathbf{M}}^{i}_{a}$ and $\hat{\mathbf{M}}^{i}_{p}$ are the estimated amplitude and phase of the speech mixture, respectively. The learned representation is exploited to modify the loss functions and learn a shared latent space between the clean speech and mixture representations. Benefited from the pre-tasks, a mapping from the mixture to the target speech is learned with the latent representation of the clean speech signal. Different from the conventional SSL methods which only estimate the magnitude, $D_{21}$ and $D_{22}$ are trained to produce both the amplitude and phase of the estimated mixture as the downstream task. Hence, using the magnitude spectrum of the speech signal as the training target can increase the accuracy of the mapping representation and improve speech enhancement performance.
\subsection{Complex-Cycle-Consistent}
\begin{figure}[htbp!]
\centering
\includegraphics[width=7cm, height=7cm]{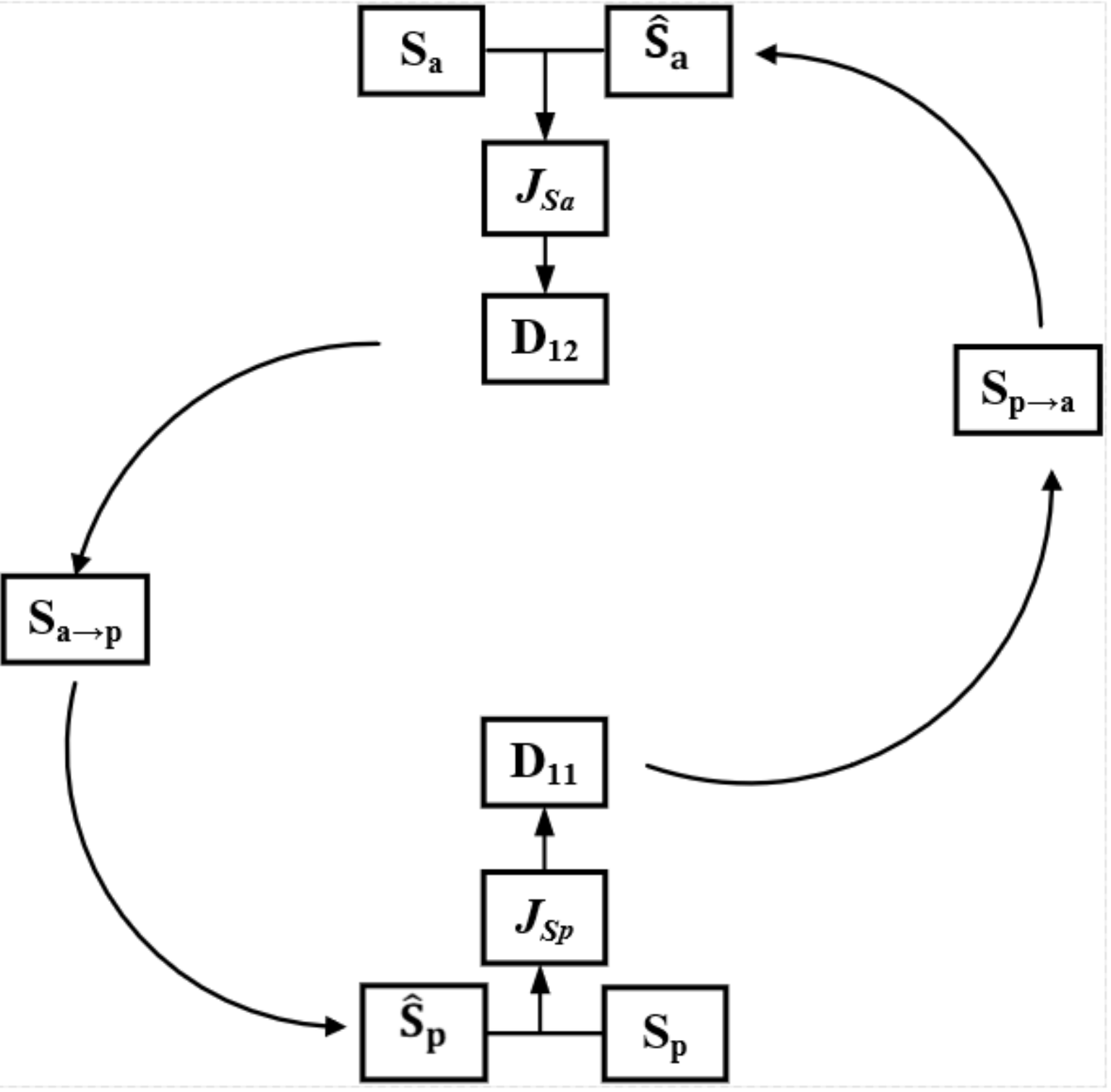}
\caption{The proposed complex-cycle-consistent for speech (CCC-S) mechanism. In the training stage, the initial reconstructions of the target speech spectra are fed into the CCC-S module. Then, the backward cycle (BC) of the phase is estimated from the amplitude domain. Consequently, the phase loss is updated with both BC and original estimations, and used to estimate the latest phase reconstruction. Similarly, the BC amplitude is estimated by using the renewed phase reconstruction and updates the combined amplitude loss. }\centering
\end{figure}

The proposed complex-cycle-consistent for speech (CCC-S) mechanism is shown in Fig. 2. As the input, the amplitude and phase of estimated speech are fed into the CCC-S module with the clean speech. A cycle-consistent constraint is exploited to minimize the reconstruction loss and further train the two autoencoders. First, the amplitude loss is estimated as equation (3), which is used to re-estimate a backward cycle (BC) of spectra from $D_{12}$. We define $a\!\rightarrow\!p$ as mapping the spectra from the amplitude to the phase and $\mathbf{S}^{i}_{a \rightarrow p}$ refers to the new phase reconstruction mapped from the amplitude loss. Then, the loss between the phase of clean speech spectra and $\mathbf{S}^{i}_{a \rightarrow p}$ is presented as:
\begin{equation}
\mathcal{J}_{\mathbf{S}_{a \rightarrow p}}= \sum_{i=1}^{I}\|\mathbf{S}^{i}_{p}-\mathbf{S}^{i}_{a \rightarrow p}\|_{2}^{2}
\end{equation}
In the training stage, the loss term $\mathcal{J}_{\mathbf{S}_{a \rightarrow p}}$ is found to perform very large compared with the loss $\mathcal{J}_{\mathbf{S}_{p}}$. Therefore, we add a constant $\theta_{1}$ and empirically set to 0.001 to constraint $\mathcal{J}_{\mathbf{S}_{a \rightarrow p}}$ and the combined phase loss can be shown as:
\begin{equation}
\mathcal{J}_{\mathbf{S}_{p}} = \mathcal{J}_{\mathbf{S}_{p}} + \theta_{1}\cdot\mathcal{J}_{\mathbf{S}_{a \rightarrow p}}
\end{equation}
The combined phase loss $\mathcal{J}_{\mathbf{S}_{p}}$ is applied to train $D_{12}$ and mapping the updated $\mathbf{S}_{a \rightarrow p}$ prepared for the next epoch. Similarly, we define $p\!\rightarrow\!a$ as mapping the spectra from the phase domain to the amplitude domain. Consequently, the $D_{11}$ obtains the combined phase loss and produces a new amplitude reconstructions as $\mathbf{S}^{i}_{p \rightarrow a}$. Thus, the loss between the amplitude of clean speech spectra and the new reconstruction as:
\begin{equation}
\mathcal{J}_{\mathbf{S}_{p \rightarrow a}}= \sum_{i=1}^{I}\|\mathbf{S}^{i}_{a}-\mathbf{S}^{i}_{p \rightarrow a}\|_{2}^{2}
\end{equation}
Accordingly, the combined amplitude loss is presented as:
\begin{equation}
\mathcal{J}_{\mathbf{S}_{a}} = \mathcal{J}_{\mathbf{S}_{a}} + \theta_{1}\cdot\mathcal{J}_{\mathbf{S}_{p \rightarrow a}}
\end{equation}
Then, the combined amplitude loss $\mathcal{J}_{\mathbf{S}_{a}}$ is applied to train $D_{11}$ and mapping the updated $\mathbf{S}_{p \rightarrow a}$ in the next epoch. The decoders are trained with the cycle-consistent and finally outputs the amplitude and phase of estimated speech spectra. The pseudocode of the proposed CCC-S module is summarized as Algorithm 1. Similarly, the DAE are trained with the BC amplitude and phase of estimated mixture spectra $\mathbf{M}_{a \rightarrow p}$ and $\mathbf{M}_{p \rightarrow a}$, respectively.

\makeatletter
\newcommand{\AlgoResetCount}{\renewcommand{\@ResetCounterIfNeeded}{\setcounter{AlgoLine}{0}}}
\newcommand{\AlgoNoResetCount}{\renewcommand{\@ResetCounterIfNeeded}{}}
\SetAlgoNoLine
\LinesNumberedHidden
\makeatother

\begin{algorithm}

  
  \SetKwInOut{Input}{input}\SetKwInOut{Output}{output}

  \Input{Amplitude of Clean speech spectra $\mathbf{S}_{a}^{i}$, Phase of Clean speech spectra $\mathbf{S}_{p}^{i}$, $i$ index of the multi-resolution feature map, learning rate $\eta$, epoch $E_{max}$}
  \Output{Estimated speech $\hat{\mathbf{S}}_{a}$ and $\hat{\mathbf{S}}_{p}$}
  \BlankLine
  \While{$E = 1$}{
  Estimate $\hat{\mathbf{S}}_{a}^{i}$ and $\hat{\mathbf{S}}_{p}^{i}$\;
  Calculate the losses: $\mathcal{J}_{\mathbf{S}_{a}}$ and $\mathcal{J}_{\mathbf{S}_{p}}$\;
  }
  \For{$E = 2, ...,$  $E_{max}$}{
   Estimate the BC $\mathbf{S}_{a \rightarrow p}$ by using $D_{12}$\;
   Update $\mathcal{J}_{\mathbf{S}_{p}}$ as equation 8\;
   Estimate the BC $\mathbf{S}_{p \rightarrow a}$ by using $D_{11}$\;
   Update $\mathcal{J}_{\mathbf{S}_{a}}$ as equation 10\;
   
     Train $D_{11}$ and $D_{12}$ by minimizing $\mathcal{J}_{\mathbf{S}_{a}}$ and $\mathcal{J}_{\mathbf{S}_{p}}$\;
 }
   Estimate $\hat{\mathbf{S}}_{a}$ and $\hat{\mathbf{S}}_{p}$ with trained $D_{11}$ and $D_{12}$\;
  \caption{Proposed complex-cycle-consistent for speech (CCC-S).}\label{algo_disjdecomp}
\end{algorithm}
\subsection{Multi-Resolution Spectra Losses}
As aforementioned, Therefore, different from the conventional SSL methods, the proposed method exploits multi-resolution feature maps as the input and the output in the encoders and decoders, respectively. Inspired by \cite{mrg}, we use the multi-resolution STFT loss as an auxiliary loss to improve the stability and efficiency of the model training. The feature map is rescaled with the same frame shift as 32 but different window sizes as 1024, 512, 256, and 128. Each STFT loss estimates the frame-level difference between the clean speech spectrogram and the corresponding reconstruction. 

We conduct three loss terms to calculate the overall loss as:
\begin{equation}
\mathcal{J}_{\text {FAE}}=\theta_{2}  \cdot \mathcal{J}_{\text {KL-FAE }}+\mathcal{J}_{\mathbf{S}}+\mathcal{J}_{\text {cyc }}
\end{equation} 
The first one is the Kullback-Leibler (KL) loss $\mathcal{J}_{\text {KL-FAE }}$ and is applied to train the latent representation $\mathbf{Z}$ closed to a normal distribution \cite{ssl}. Then, $\theta_{2}$ is the coefficient of $\mathcal{J}_{\text {KL-FAE }}$ and empirically set to 0.001. The $\mathcal{J}_{\mathbf{S}}$ denotes the sum of amplitude and phase with four multi-resolution losses between the clean speech feature and the corresponding reconstruction as:
\begin{equation}
\mathcal{J}_{\text {S}}= \sum_{i=1}^{I}(\|\mathbf{S}^{i}_{a}-\hat{\mathbf{S}}^{i}_{a}\|_{2}^{2}+\|\mathbf{S}^{i}_{p}-\hat{\mathbf{S}}^{i}_{p}\|_{2}^{2})
\end{equation}

Similarly, the cycle loss $\mathcal{J}_{\mathrm{cyc}}$ consists of $\mathcal{J}_{\mathbf{S}}$ and the loss between the latent representation and the corresponding reconstruction:
\begin{equation}
\mathcal{J}_{\mathrm{cyc}}=\mathcal{J}_{\mathbf{S}}+\theta_{3} \cdot\sum_{i=1}^{I}\left\|\mathbf{Z}^{i}-\hat{\mathbf{Z}}^{i}\right\|_{2}^{2}
\end{equation}
where $\mathbf{Z}^{i}$ and $\hat{\mathbf{Z}}^{i}$ are the clean and estimated representation of the target speech signal at the $i$-th multi-resolution feature map, respectively. Moreover, $\theta_{3}$ is the coefficient of representation loss and empirically set to 0.001. The pseudocode of the proposed phase-aware multi-resolution FAE is summarized as Algorithm 2. Similarly, the loss for the speech mixture is calculated.

\makeatletter
\SetAlgoNoLine
\LinesNumberedHidden
\makeatother

\begin{algorithm}

  
  \SetKwInOut{Input}{input}\SetKwInOut{Output}{output}

  \Input{Clean spectra $\mathbf{S}^{i}$, $i$ index of the multi-resolution feature map, learning rate $\eta$, epoch $E_{max}$}
  \Output{Estimated clean speech $\hat{\mathbf{S}}$}
  \BlankLine
  Initialize FAE parameters\;
  \For{$E = 1, 2, ...,$  $E_{max}$}{
   
    \For{$i = 1\leftarrow4$}{
       Obtain $\mathbf{Z}^{i}$ $\leftarrow$ $E_{1}(\mathbf{S}^{i})$\; 
       // Train the Encoder with multi-resolution spectra\;  
      }
       
    Learn the clean speech representation $\mathbf{Z}^{i}$\;
    Amplitudes $\mathbf{S}_{a}$ and $\mathbf{Z}_{a}$ are fed into $D_{11}$\;
    Estimate $\hat{\mathbf{Z}}_{a}^{1}$, $\hat{\mathbf{S}}_{a}^{1}$ $\leftarrow$ $D_{11}$ ($\mathbf{Z}_{a}^{1}$, $\mathbf{S}_{a}^{1}$)\;
    \For{$i = 2, 3, 4$}{
      Estimate $\hat{\mathbf{Z}}_{a}^{i}$, $\hat{\mathbf{S}}_{a}^{i}$ $\leftarrow$ $D_{11}$ ($\mathbf{Z}_{a}^{i-1}$, $\mathbf{S}_{a}^{i-1}$)\;
      // Obtain the estimated amplitude of multi-resolution spectra\;
    }
   Phases $\mathbf{S}_{p}$ and $\mathbf{Z}_{p}$ are fed into $D_{12}$\;
   Estimate $\hat{\mathbf{Z}}_{p}^{1}$, $\hat{\mathbf{S}}_{p}^{1}$ $\leftarrow$ $D_{11}$ ($\mathbf{Z}_{p}^{1}$, $\mathbf{S}_{p}^{1}$)\;
    \For{$i = 2, 3, 4$}{
      Estimate $\hat{\mathbf{Z}}_{p}^{i}$, $\hat{\mathbf{S}}_{p}^{i}$ $\leftarrow$ $D_{11}$ ($\mathbf{Z}_{p}^{i-1}$, $\mathbf{S}_{p}^{i-1}$)\;
      // Obtain the estimated phase of multi-resolution spectra\;
    }
    $\hat{\mathbf{S}}=\hat{\mathbf{S}}_{a} \cdot e^{j \hat{\mathbf{S}}_{p}}$\;
    Train FAE by minimizing $\mathcal{L}_{\text {FAE}}$, $\mathcal{J}_{\mathbf{S}_{a}^{i}}$, and $\mathcal{J}_{\mathbf{S}_{p}^{i}}$\;
 }
  \caption{FAE Pipeline.}\label{algo_disjdecomp}
\end{algorithm}

\begin{table*}[htbp!]
\caption{averaged speech enhancement performance in terms of three noise interferences at four SNR levels in ipad$\char`_$livingroom1.}
\centering
\begin{tabular}{ccccccccccccccccc}
\hline
& \multicolumn{4}{c}{PESQ} & %
    \multicolumn{4}{c}{CSIG} & \multicolumn{4}{c}{CBAK}& \multicolumn{4}{c}{COVL}\\
\cline{2-17}
SNR (dB)  &-10&-5 & 0 & 5 &-10 &-5 & 0 & 5 &-10 &-5 & 0 & 5 &-10 &-5 & 0 & 5 \\
CL \cite{cl}   &1.43 &1.52&1.54&1.60 &1.96&2.20&2.30&2.40 & 1.57& 1.76&1.92&2.03&1.55&1.77&1.86&1.94 \\
 SSE \cite{ssl}   &1.48&1.53&1.56&1.58 &2.04&2.30&2.39&2.45 & 1.63& 1.83&1.94&2.10&1.68&1.81&1.88&2.00 \\
PT-FT \cite{ssl1} &1.52&1.55&1.59&1.62 &2.10&2.28&2.34&2.43 & 1.67& 1.81&1.96&2.08&1.68&1.78&1.89&2.00 \\
\textit{Proposed}   &{\bfseries 1.67}&{\bfseries 1.73}&{\bfseries 1.78}&{\bfseries 1.80}&{\bfseries 2.41} &{\bfseries 2.47}&{\bfseries 2.51}&{\bfseries 2.47} & {\bfseries 1.87}& {\bfseries 1.94}& {\bfseries 2.12}& {\bfseries 2.26}& {\bfseries 1.82}& {\bfseries 1.90}& {\bfseries 1.99}& {\bfseries 2.06}\\
 \hline 
\end{tabular}
\end{table*}

\begin{table*}[htbp!]
\caption{averaged speech enhancement performance in terms of three noise interferences at four SNR levels in ipad$\char`_$bedroom1.}
\centering
\begin{tabular}{ccccccccccccccccc}
\hline
& \multicolumn{4}{c}{PESQ} & %
    \multicolumn{4}{c}{CSIG} & \multicolumn{4}{c}{CBAK}& \multicolumn{4}{c}{COVL}\\
\cline{2-17}
SNR (dB)  &-10&-5 & 0 & 5 &-10 &-5 & 0 & 5 &-10 &-5 & 0 & 5 &-10 &-5 & 0 & 5 \\
CL \cite{cl}   &1.45 &1.57&1.59&1.61&1.93 &2.25&2.32&2.39& 1.69 & 1.82&1.99&2.08&1.70&1.82&1.90&2.03  \\
 SSE \cite{ssl}  &1.50&1.59&1.62&1.65&2.11 &2.34&2.43&2.49& 1.72 & 1.88&1.97&2.16&1.73&1.84&1.89&2.02 \\
PT-FT \cite{ssl1}  &1.57&1.64&1.73&1.74 &2.16&2.33&2.46&2.51 & 1.75& 1.91&2.03&2.19&1.77&1.85&1.94&2.05 \\
\textit{Proposed}   &{\bfseries 1.79}&{\bfseries 1.88}&{\bfseries 1.92}&{\bfseries 1.95}&{\bfseries 2.40} &{\bfseries 2.49}&{\bfseries 2.56}&{\bfseries 2.61} & {\bfseries 1.94}& {\bfseries 2.02}& {\bfseries 2.16}& {\bfseries 2.25}& {\bfseries 1.88}& {\bfseries 1.96}& {\bfseries 2.05}& {\bfseries 2.13}\\
 \hline 
\end{tabular}
\end{table*}

\begin{table*}[htbp!]
\caption{averaged speech enhancement performance in terms of three noise interferences at four SNR levels in ipad$\char`_$confroom1.}
\centering
\begin{tabular}{ccccccccccccccccc}
\hline
& \multicolumn{4}{c}{PESQ} & %
    \multicolumn{4}{c}{CSIG} & \multicolumn{4}{c}{CBAK}& \multicolumn{4}{c}{COVL}\\
\cline{2-17}
SNR (dB)  &-10&-5 & 0 & 5 &-10 &-5 & 0 & 5 &-10 &-5 & 0 & 5 &-10 &-5 & 0 & 5 \\
 CL \cite{cl} &1.48&1.58&1.62&1.63 &2.09&2.26&2.33&2.44 & 1.77& 1.84&2.00&2.09&1.81&1.85&1.92&2.06  \\
 SSE \cite{ssl}  &1.53&1.61&1.65&1.66 &2.12&2.35&2.46&2.47 & 1.78& 1.93&2.00&2.17&1.80&1.85&1.90&2.05 \\
PT-FT \cite{ssl1}   &1.60&1.66&1.74&1.77 &2.18&2.34&2.45&2.53 & 1.83& 1.94&2.05&2.23&1.96&\textbf{2.02}&{\bfseries 2.07}&2.10 \\
\textit{Proposed}   &{\bfseries 1.81}&{\bfseries 1.89}&{\bfseries 1.96}&{\bfseries 1.98}&{\bfseries 2.36} &{\bfseries 2.50}&{\bfseries 2.57}&{\bfseries 2.65} & {\bfseries 1.99}& {\bfseries 2.04}& {\bfseries 2.17}& {\bfseries 2.25}& {\bfseries 1.97}& 2.00& {\bfseries 2.07}& {\bfseries 2.19}\\

 \hline 
\end{tabular}
\end{table*}
\section{EXPERIMENTAL RESULTS}
\subsection{Datasets and Comparisons}
In the training stage, 600 clean utterances from 20 different speakers with three room environments are randomly selected from the DAPS dataset \cite{daps}. The training data consists of 10 male and 10 female speakers each reading out 5 utterances and recorded in different indoor environments with different real room impulse responses (RIRs). In each environment, we first randomly select 12 utterances to generate the training data to train the FAE. Then, the rest 188 utterances are exploited for DAE to obtain the estimated mixtures. Therefore, the training data in the FAE and DAE is unseen and not overlapping. Moreover, we use three background noises ($factory$, $babble$, and $cafe$) from the NOISEX dataset \cite{noise} and four SNR levels (-10, -5, 0, and 5 dB) to generate the mixtures. It is highlighted that the training data used in the training stage is unpaired. In the testing stage, 300 clean utterances of 10 speakers are randomly selected and used to generate the mixtures with the same background noises and SNR levels as the configuration in training stage. 

We compare the proposed method with three recent SSL speech enhancement approaches \cite{ssl, cl, ssl1} on two publicly-available datasets. The first method is SSE \cite{ssl} which exploits two autoencoders to process pre-task and downstream task, respectively. Different from the proposed method, the speech mixture in \cite{ssl} is generated with the clean speech signal and the corresponding reverberation. The second method is pre-training fine-tune (PT-FT) \cite{ssl1}, which uses three models and three SSL approaches for pre-training: speech enhancement, masked acoustic model with alteration (MAMA) used in TERA \cite{mama} and continuous contrastive task (CC) used in wav2vec 2.0 \cite{wav}. We reproduce the PT-FT method with DPTNet model \cite{DT} and three pre-tasks because it shows the best enhancement performance in \cite{ssl1}. The third method applies a simple contrastive learning (CL) procedure which treats the abundant noisy data as makeshift training targets through pairwise noise injection \cite{ssl5}. In the baseline, the recurrent neural network (RNN) outputs with a fully-connected dense layer with sigmoid activation to estimate a time-frequency mask which is applied onto the noisy speech spectra. The configuration difference as below.

\begin{table}[htbp!]
\centering
\caption{Comparison of SSL methods with the proposed approach. The PT-FT method use 50,800 paired utterances in the training stage, however, only 200 unpaired utterances are applied in the proposed method. Moreover, the number of pre-tasks is set to 3 and 2 in the PT-FT and proposed method, respectively.}
\small\addtolength{\tabcolsep}{-1pt}
\begin{tabular}{c|c|c|c}
\hline
 & SSE (2020)  & PT-FT (2021) &\textit{Proposed } \\    
 \hline
Noise & \xmark & \checkmark  & \checkmark    \\
\hline
Paired Data & \xmark & \checkmark & \xmark    \\
\hline
  Multiple Models   &\checkmark&\xmark & \checkmark\\
 \hline 
  Single Pre-Task  &\checkmark &\xmark &\xmark\\
 \hline 
\end{tabular}
\end{table}
\subsection{Experimental Setup}
The proposed method is trained by using the Adam optimizer with a learning rate of 0.001 and the batch size is 20. The number of training epochs for FAE and DAE are 700 and 1500, respectively. All the experiments are run on a work station with four Nvidia GTX 1080 GPUs and 16 GB of RAM. The complex spectra have 513 frequency bins for each frame as a Hanning window and a discrete Fourier transform (DFT) size of 1024 samples are applied.

\begin{figure*}[htbp!]
\centering
\includegraphics[width=17.5cm, height=4.5cm]{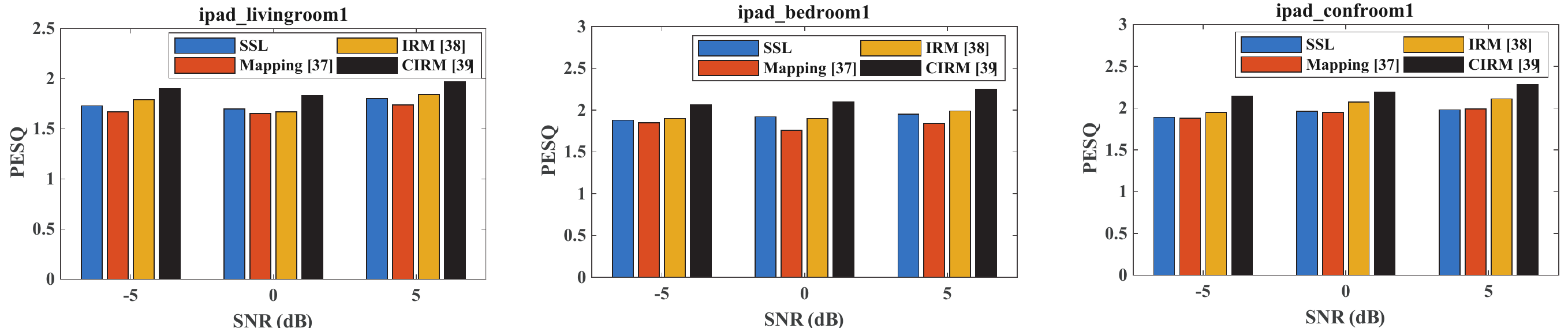}
\caption{Comparisons with supervised methods in three environments (ipad$\char`_$livingroom1, ipad$\char`_$bedroom1, and ipad$\char`_$confroom1). Each result is the average
value of 300 experiments in terms of three SNR levels (-5, 0, and 5 dB).}\centering
\end{figure*}

According to \cite{ssl}, we use composite metrics that approximate the Mean Opinion Score (MOS) including COVL: MOS predictor of overall signal quality, CBAK: MOS predictor of background-noise intrusiveness, CSIG: MOS predictor of signal distortion \cite{mos} and Perceptual Evaluation of Speech Quality (PESQ). Besides, signal-to-distortion ratio (SDR) is evaluated in terms of baselines and the proposed method. Higher values of the measurements imply the better enhancement performance.

\subsection{Comparison with SSL methods}
As the proposed method achieves the best speech enhancement performance with the multi-resolution and phase information, it is further compared with the state-of-the-art SSL methods in TABLES II-IV.

It can be seen from TABLES II-IV that the proposed method outperforms the state-of-the-art SSL methods in terms of all three performance measurements. The environment ipad$\char`_$livingroom1 is relatively more reverberant compared to the other two environments \cite{daps}, while the improvement is still significant. For example, in TABLE II, the proposed method has 15.6$\%$, 14.1$\%$, and 11.3$\%$ improvements compared with the CL, SSE, and PT-FT methods in terms of PESQ at 0 dB, respectively. Besides, speech enhancement comparisons at four different SNR levels are shown in tables. From the experimental results, the performance improvement compared to the baselines are obvious even at relatively low SNR level i.e., -10 dB. The proposed method has 14.7$\%$, 11.1$\%$, and 8.2$\%$ improments compared with the PT-FT method in terms of CSIG at -10 dB SNR level in three environments.

In \cite{ssl1}, the original PT-FT method is trained with Libri1Mix train-360 set \cite{lib} which contains 50,800 utterances. However, in the comparison experiments, we use the limited amount of training utterances (200). Therefore, the speech enhancement performance of the PT-FT suffers a significant degradation compared with the original implementation. The latent representation and the masking module have limitations, however, the proposed method takes advantage of both approaches and mitigates the speech enhancement problem. Thus, the speech enhancement performance is improved compared with only learning the clean speech representation in the SSE method.
\subsection{Comparison with SL methods}
Recently, most of speech enhancement methods are developed based on supervised learning (SL) due to the promising performance under the sufficient training data. However, in the practical scenarios, the training frequently suffers the problem which lacking in paired data. Therefore, in order to show the competitiveness of the proposed SSL method, the mapping- and masking-based supervised methods are reproduced with the same number of training data \cite{mapping, IRM1, cirm}. The SL baselines are implemented with deep neural networks (DNNs) which use three hidden layers, each having 1024 rectified linear hidden units as the original implementations. Apart form the ideal ratio mask (IRM), we also compare the proposed phase-aware method with the complex ideal ratio mask (cIRM). The experimental results of comparisons with the SL methods are presented in Fig. 3.




\begin{table*}[htbp!]
\caption{ablation study of three contributions}
\centering
\begin{tabular}{cccccccc}
\hline

\multicolumn{3}{c}{Ablation Settings} & %
    \multirow{2}{*}{PESQ}     & \multirow{2}{*}{CSIG}& \multirow{2}{*}{CBAK}& \multirow{2}{*}{COVL}& \multirow{2}{*}{SDR (dB)}\\
\cline{1-3}
  Multi-Resolution&Phase-Aware& CCC \\
 \hline
\xmark &\xmark &\xmark & 1.48&2.28&1.90&1.84 &4.76 \\
\checkmark   &\xmark &\xmark   &1.56 & 2.39 & 1.94&1.88 &5.16  \\
\xmark  &  \checkmark &\xmark  &1.58 & 2.41 & 1.97&1.84 &5.41 \\
\xmark  &  \checkmark &\checkmark  &1.73 & 2.44 & 2.11&1.98 &7.02 \\
\checkmark   &\checkmark  &\xmark &1.69 & 2.42 & 2.09&1.94 &6.73 \\
\checkmark   &\checkmark   &\checkmark&{\bfseries 1.77} &{\bfseries 2.47} & {\bfseries 2.12}& {\bfseries 2.01}& {\bfseries 8.93}\\
 \hline 
\end{tabular}
\end{table*}
From Fig. 3, we can observe that the proposed SSL method shows better performance than the mapping-based method while performs limited compared with the masking-based methods. On the one hand, the compared baselines are not state-of-the-art approaches. However, the SSL research in speech enhancement problem just started \cite{ssl}. We simply provide the comparison between the SSL and SL study to show the competitiveness of the proposed method. Besides, the experiments are set up in a challenging environments with high reverberation as the practical scenarios. Therefore, the improvements of the proposed and baselines are relatively limited.

\subsection{Ablation Study}
We investigate the effectiveness of each contribution based on the DAPS dataset. In the baseline, a single spectrogram is obtained from the output layer of the decoders in FAE and DAE to compare to the proposed multi-resolution method. The experimental results in terms of four performance measurements are shown in TABLE VI. Due to the dependency between the phase-aware and CCC module, the ablation experiments with the CCC module but without the phase-aware are not conducted.

Initially, the effectiveness of the multi-resolution spectra losses is studied. We conduct two sets of experiments that differs at the resolutions of input spectra. First, the single-resolution spectra are fed into the encoder. Then, the proposed method has a PESQ improvement of 0.08 after the multi-resolution spectra losses are introduced. As for the reason, the proposed method utilizes the combination of multi-resolution spectra losses to learn multiple levels of acoustic properties in a balanced way. Consequently, different information distributed in the multi-resolution feature maps is extracted to improve the accuracy of the target speech estimation.


%
Moreover, the experiment is conducted by adding the proposed phase-aware decoders. From TABLE VI, it can be observed that the performance is significantly improved by the proposed phase-aware method among all four measurements. For example, in terms of PESQ, the performance is improved from 1.56 to 1.69, which further confirms that the proposed method with the phase-aware decoders can boost the enhancement performance. In the baselines, the speech signal is reconstructed by using the noisy phase and the estimated magnitude, which causes a phase loss between the clean speech signal and the corresponding reconstruction. However, the proposed phase-aware method utilizes $D_{12}$ and $D_{22}$ to estimate the phase of the target speech signal and speech mixture, respectively, and improve the accuracy of estimation.

Furthermore, the CCC-S and CCC-M modules are added to the FAE and DAE, respectively. Compared with three baselines, the proposed CCC method brings a obvious improvement in terms of all performance measurements. For instance, the proposed method has a COVL improvement of 0.07 after the CCC modules are introduced. In SSL study, due to the limited training data, the potential linking information between the amplitude and phase plays an important role in the speech enhancement problem. With the proposed CCC method, each of the amplitude and phase is estimated with the updated reconstruction of the other and the desired speech information is better preserved in the enhanced features.


\section{CONCLUSIONS}

In this paper, we proposed a self-supervised learning based method with the complex spectrogram to address the monaural speech enhancement problem. Different from the previous single task in SSL, our proposed method contained two pre-tasks, which could help to estimate both amplitude and phase information of the desired speech signal. Besides, the proposed complex-cycle-consistent mechanism provided mappings between the amplitude and phase to update the combined losses and further refine the estimation accuracy. In order to better utilize information from the desired speech signal, we proposed a multi-resolution spectra losses based on the multi-resolution feature maps. The experimental results showed that the proposed method outperforms the state-of-the-art SSL approaches. 

In the future work, the performance of the proposed method can be further improved by decoupling magnitude and phase estimation \cite{qqk}. Finally, the proposed method denoises the speech mixture in a highly reverberant environment. Future work should be dedicated to exploit the dereverberation pre-task \cite{IET, SIVA} to further refine the speech enhancement performance.

\ifCLASSOPTIONcaptionsoff
  \newpage
\fi



\bibliographystyle{IEEEtran}
%
\bibliography{Special issue from AAAI workshop (Copy)}

\begin{thebibliography}{10}
\providecommand{\url}[1]{#1}
\csname url@samestyle\endcsname
\providecommand{\newblock}{\relax}
\providecommand{\bibinfo}[2]{#2}
\providecommand{\BIBentrySTDinterwordspacing}{\spaceskip=0pt\relax}
\providecommand{\BIBentryALTinterwordstretchfactor}{4}
\providecommand{\BIBentryALTinterwordspacing}{\spaceskip=\fontdimen2\font plus
\BIBentryALTinterwordstretchfactor\fontdimen3\font minus
  \fontdimen4\font\relax}
\providecommand{\BIBforeignlanguage}[2]{{%
\expandafter\ifx\csname l@#1\endcsname\relax
\typeout{** WARNING: IEEEtran.bst: No hyphenation pattern has been}%
\typeout{** loaded for the language `#1'. Using the pattern for}%
\typeout{** the default language instead.}%
\else
\language=\csname l@#1\endcsname
\fi
#2}}
\providecommand{\BIBdecl}{\relax}
\BIBdecl

\bibitem{yiluo}
Y.~Luo, C.~Han, and N.~Mesgarani, ``{Ultra-lightweight speech separation via
  group communication},'' \emph{IEEE International Conference on Acoustics,
  Speech and Signal Processing (ICASSP)}, 2021.

\bibitem{asr}
X.~K. Chang, T.~Maekaku, P.~C. Guo, J.~Shi, Y.-J. Lu, A.~S. Subramanian, T.~Z.
  Wang, S.-W. Yang, Y.~Tsao, H.-Y. Lee, and S.~Watanabe, ``An exploration of
  self-supervised pretrained representations for end-to-end speech
  recognition,'' \emph{arXiv preprint arXiv:2110.04590}, 2021.

\bibitem{ssl6}
L.~Jing, P.~Vincent, Y.~LeCun, and Y.~D. Tian, ``Understanding dimensional
  collapse in contrastive self-supervised learning,'' \emph{arXiv preprint
  arXiv:2110.09348}, 2021.

\bibitem{ssl}
Y.-C. Wang, S.~Venkataramani, and P.~Smaragdis, ``{Self-supervised learning for
  speech enhancement},'' \emph{International Conference on Machine Learning
  (ICML)}, 2020.

\bibitem{ssl2}
Z.~H. Du, M.~Lei, J.~Q. Han, and S.~L. Zhang, ``{Self-supervised adversarial
  multi-task learning for vocoder-based monaural speech enhancement},''
  \emph{Interspeech}, 2020.

\bibitem{csse}
Z.~Meng, J.~Y. Li, Y.~F. Gong, and B.-H. Juang, ``{Cycle-consistent speech
  enhancement},'' \emph{Interspeech}, 2018.

\bibitem{TCN}
G.~C. Yu, Y.~T. Wang, H.~Wang, Q.~Zhang, and C.~S. Zheng, ``{A two-stage
  complex network using cycle-consistent generative adversarial networks for
  speech enhancement},'' \emph{Speech Communication}, vol. 134, pp. 42 -- 54,
  2021.

\bibitem{multis}
W.~Liu, M.~Sun, X.~Zhang, and T.~F.~Z. H.~Van~Hamme, ``{A multi-resolution
  front-end for end-to-end speech anti-spoofing},'' \emph{arXiv preprint
  arXiv:2110.05087}, 2021.

\bibitem{twos}
J.~Li, X.~R. Xie, N.~Yan, and L.~Wang, ``{Two streams and two resolution
  spectrograms model for end-to-end automatic speech recognition},''
  \emph{arXiv preprint arXiv:2108.07980}, 2021.

\bibitem{icme}
X.~Y. Ma, T.~Y. Liang, S.~S. Zhang, S.~Huang, and L.~He, ``{Improved lightcnn
  with attention modules for ASV spoofing detection},'' \emph{IEEE
  International Conference on Multimedia and Expo (ICME)}, 2021.

\bibitem{weak}
Q.~Q. Kong, H.~H. Liu, X.~J. Du, L.~Chen, R.~Xia, and Y.~X. Wang, ``{Speech
  enhancement with weakly labelled data from AudioSet},'' \emph{Interspeech},
  2021.

\bibitem{poco}
U.~Isik, R.~Giri, N.~Phansalkar, J.-M. Valin, K.~Helwani, and A.~Krishnaswamy,
  ``{PoCoNet: better speech enhancement with frequency-positional embeddings,
  semi-supervised conversational data, and biased loss},'' \emph{Interspeech},
  2020.

\bibitem{semi}
S.~Seki, M.~Takada, and T.~Toda, ``{Semi-supervised self-produced speech
  enhancement and suppression based on joint source modeling of air- and
  body-conducted signals using variational autoencoder},'' \emph{Interspeech},
  2020.

\bibitem{ssl7}
Y.-C. Chen, S.-W. Yang, C.-K. Lee, S.~See, and H.-Y. Lee, ``Speech
  representation learning through self-supervised pretraining and multi-task
  finetuning,'' \emph{arXiv preprint arXiv:2110.09930}, 2021.

\bibitem{CSA1}
Y.~Sun, Y.~Xian, W.~Wang, and S.~M. Naqvi, ``Monaural source separation in
  complex domain with long short-term memory neural network','' \emph{IEEE
  Journal of Selected Topics in Signal Processing}, vol.~13, no.~2, pp. 359 --
  369, 2019.

\bibitem{mcgn}
Y.~Xian, Y.~Sun, W.~W. Wang, and S.~M. Naqvi, ``A multi-scale feature
  recalibration network for end-to-end single channel speech enhancement,''
  \emph{IEEE Journal of Selected Topics in Signal Processing}, vol.~15, no.~1,
  pp. 143--155, 2021.

\bibitem{mr}
Z.~C. Peng, J.~W. Dang, M.~Unoki, and M.~Akagi, ``Multi-resolution
  modulation-filtered cochleagram feature for lstm-based dimensional emotion
  recognition from speech,'' \emph{Neural Networks}, vol. 140, pp. 261--273,
  2021.

\bibitem{pwgan}
R.~Yamamoto, E.~Song, and J.-M. Kim, ``{Parallel waveGAN: a fast waveform
  generation model based on generative adversarial networks with
  multi-resolution spectrogram},'' \emph{IEEE International Conference on
  Acoustics, Speech and Signal Processing (ICASSP)}, 2020.

\bibitem{vogan}
J.~Yang, J.~Lee, Y.~Kim, H.~Cho, and I.~Kim, ``{VocGAN: a high-fidelity
  real-time vocoder with a hierarchically-nested adversarial network},''
  \emph{Interspeech}, 2020.

\bibitem{ganno1}
M.~Y. Li, J.~Lin, Y.~Y. Ding, Z.~J. Liu, J.-Y. Zhu, and S.~Han, ``{GAN
  compression: efficient architectures for interactive conditional GANs},''
  \emph{Conference on Computer Vision and Pattern Recognition (CVPR)}, 2020.

\bibitem{ganno}
Y.~Tian, X.~Peng, L.~Zhao, S.~T. Zhang, and D.~N. Metaxas, ``Cr-gan: learning
  complete representations for multi-view generation,'' \emph{International
  Joint Conference on Artificial Intelligence (IJCAI)}, 2018.

\bibitem{ganno2}
Z.~Ding, Y.~F. Xu, W.~J. Xu, G.~Parmar, Y.~Yang, M.~Welling, and Z.~W. Tu;,
  ``Guided variational autoencoder for disentanglement learning,''
  \emph{Conference on Computer Vision and Pattern Recognition (CVPR)}, 2020.

\bibitem{VCAE}
D.~T. Braithwaite and W.~B. Kleijn, ``Speech enhancement with variance
  constrained autoencoders,'' \emph{Interspeech}, 2019.

\bibitem{TAI}
Y.~Li, Y.~Sun, K.~Horoshenkov, and S.~M. Naqvi, ``{Domain adaptation and
  autoencoder based unsupervised speech snhancement},'' \emph{IEEE Transactions
  on Artificial Intelligence}, 2021.

\bibitem{letter}
Z.-Q. Wang, G.~Wichern, and J.~L. Roux, ``On the compensation between magnitude
  and phase in speech separation,'' \emph{IEEE Signal Processing Letters},
  2021.

\bibitem{ssl5}
A.~Sivaraman, S.~Kim, and M.~Kim, ``Personalized speech enhancement through
  self-supervised data augmentation and purification,'' \emph{Interspeech},
  2021.

\bibitem{mrg}
H.~Y. Kim, J.~Yoon, S.~J. Cheon, W.~H. Kang, and N.~Kim, ``A multi-resolution
  approach to gan-based speech enhancement,'' \emph{Applied Sciences}, vol.~11,
  no.~2, p. 721, 2021.

\bibitem{cl}
A.~Sivaraman and M.~Kim, ``Self-supervised learning from contrastive mixtures
  for personalized speech enhancement,'' \emph{arXiv preprint
  arXiv:2011.03426}, 2020.

\bibitem{ssl1}
S.-F. Huang, S.-P. Chuang, D.-R. Liu, Y.-C. Chen, G.-P. Yang, and H.-Y. Lee,
  ``{Stabilizing label assignment for speech separation by self-supervised
  pre-training},'' \emph{Interspeech}, 2021.

\bibitem{daps}
G.~J. Mysore, ``{Can we automatically transform speech recorded on common
  consumer devices in real-world environments into professional production
  quality speech?—a dataset, insights, and challenges},'' \emph{IEEE Signal
  Processing Letters}, vol.~22, no.~8, pp. 1006 -- 1010, 2014.

\bibitem{noise}
A.~Varga and H.~J.~M. Steeneken, ``Assessment for automatic speech recognition:
  Ii. noisex-92: a database and an experiment to study the effect of additive
  noise on speech recognition systems,'' \emph{IEEE Transactions on Audio,
  Speech, and Language Processing}, vol.~12, no.~3, pp. 247 -- 251, 1993.

\bibitem{mama}
A.~T. Liu, S.-W. Li, and H.-Y. Lee, ``{TERA: self-supervised learning of
  transformer encoder representation for speech},'' \emph{IEEE/ACM Transactions
  on Audio, Speech, and Language Processing}, vol.~29, pp. 2351 -- 2366, 2021.

\bibitem{wav}
A.~Baevski, H.~Zhou, A.~Mohamed, and M.~Auli, ``{wav2vec 2.0: a framework for
  self-supervised learning of speech representations},'' \emph{Neural
  Information Processing Systems (NeurIPS)}, 2020.

\bibitem{DT}
J.~J. Chen, Q.~R. Mao, and D.~Liu, ``{Dual-path transformer network: direct
  context-aware modeling for end-to-end monaural speech separation},''
  \emph{Interspeech}, 2020.

\bibitem{mos}
Y.~Hu and P.~C. Loizou, ``{Evaluation of objective quality measures for
  speech},'' \emph{IEEE Transactions on Audio, Speech, and Language
  Processing}, vol.~16, no.~1, pp. 229 -- 238, 2008.

\bibitem{lib}
J.~Cosentino, M.~Pariente, S.~Cornell, A.~Deleforge, and E.~Vincent,
  ``{LibriMix: an open-source dataset for generalizable speech separation},''
  \emph{Interspeech}, 2020.

\bibitem{mapping}
Y.~Xu, J.~Du, L.-R. Dai, and C.-H. Lee, ``A regression approach to speech
  enhancement based on deep neural networks,'' \emph{IEEE/ACM Transanctions on
  Audio Speech and Language Processing}, vol.~23, no.~1, pp. 7--19, 2015.

\bibitem{IRM1}
Y.~Wang, A.~Narayanan, and D.~L. Wang, ``{On training targets for supervised
  speech separation},'' \emph{IEEE/ACM Transactions on Audio, Speech, and
  Language Processing}, vol.~22, no.~12, pp. 1849--1858, 2014.

\bibitem{cirm}
D.~S. Williamson and D.~L. Wang, ``Time-frequency masking in the complex domain
  for speech dereverberation and denoising,'' \emph{IEEE/ACM Transanctions on
  Audio Speech and Language Processing}, vol.~25, no.~7, pp. 1492--1501, 2017.

\bibitem{qqk}
Q.~Q. Kong, Y.~Cao, H.~H. Liu, K.~Choi, and Y.~X. Wang, ``Decoupling magnitude
  and phase estimation with deep resunet for music source separation,''
  \emph{arXiv preprint arXiv:2109.05418}, 2021.

\bibitem{IET}
Y.~Li, Y.~Sun, and S.~M. Naqvi, ``{Single-channel dereverberation and denoising
  based on lower band trained SA-LSTMs},'' \emph{IET Signal Processing},
  vol.~14, no.~10, pp. 774 -- 782, 2021.

\bibitem{SIVA}
Y.~Sun, W.~Wang, J.~A. Chambers, and S.~M. Naqvi, ``{Enhanced time-frequency
  masking by using neural networks for monaural source separation in
  reverberant room environments},'' \emph{European Signal Processing Conference
  (EUSIPCO)}, 2018.

\end{thebibliography}
\end{document}